\documentclass[11pt, oneside,fleqn]{article}   	
\usepackage{geometry}                		
\usepackage{graphicx}				
\usepackage{amssymb}

 \usepackage{helvet} 

\usepackage{wrapfig}

\usepackage{float}

\newcommand{\vect}[1]{\mbox{$\mathrm{\textbf #1}$}}
\newcommand{\nvec}[1]{\mbox{\boldmath $\hat{#1}$}}
\newcommand{\Tens}[1]{\mbox{\fontfamily{phv}\selectfont{\normalsize{\bfseries{#1}}}}}
\newcommand{\tens}[1]{\mbox{\fontfamily{phv}\selectfont{\footnotesize{\bfseries{#1}}}}}

\title{A Physical Model for Self-Similar Seashells}
\author{Paul A. Reiser (reiser.paul@gmail.com)}

\begin{document}
\maketitle
\tableofcontents
\newpage

\section{Abstract}

This paper presents a simple physical model for self-similar (gnomonic, or first-order) seashell growth which is expressed in coordinate-free terms. The shell is expressed as the solution of a differential equation which expresses the growth dynamics, and may be used to investigate shell growth from both the local viewpoint of the organism building it and moving with the shell opening (aperture), as well as that of a researcher making global measurements upon a complete motionless shell. Coordinate systems needed to express the global and local descriptions of the shell are chosen. The parameters of growth, or their information equivalent, remain constant in the local system, and are used by the organism to build the shell, and are likely mirrored in the DNA of the organism building it. The transformations between local and global representations are provided. The global model of Cortie\cite{Cortie}, which is very similar to the present model, is expressed in terms of the present model, and the global parameters provided by Cortie for various species of mollusk may be used to calculate the equivalent local parameters.Mathematica code is provided to implement these transformations, as well as to plot the shells using both global and local parameters.

\section{Introduction}

Seashells of almost all types exhibit certain striking qualities and similarities. In particular, they approximate to varying degrees the property of self-similarity\footnote{In the present context, self-similarity means that the description of the shell at time $t+\Delta t$ is simply a magnified and rotated version of the shell at time $t$.}: Beyond a certain point in time, younger individuals practically differ from older individuals only in scale. Furthermore, to varying degrees, seashells appear to be different variations on one general theme: They wrap themselves in what is known as a logarithmic helico-spiral, which is self-similar. These properties suggest that a self-similar seashell can be described mathematically with a relatively small number of parameters which will account for the various shapes of the shells.

Models of seashell shapes generally describe the shell in terms of parameters derived from "global" measurements of the shell (e.g., the angle of expansion, the size of the aperture, the chirality or handedness of the rotation). However, it is assumed that the organism does not have access to these global parameters but rather builds its shell using "local" parameters that simply describe how to grow the shell, given only the present aperture.

In the present model, the growth at the aperture in an infinitesimal time interval is generally represented by an infinitesimal expansion, rotation, and displacement of the present aperture, without reference to the global shell geometry and without the need for an internal clock. (This may not be true in the case of  shells with periodic knobs, such as \textit{Epitonium}.) The present model is a special case of higher order models\cite{Illert1989} and more detailed models which describe shell growth in terms of the biological and mechanical processes occurring at the aperture edge\cite{Rudraraju,Ambrosi}. This differential process of the present model, expressed as a differential equation, may then be integrated to obtain the shape of the entire shell. It is desired to express the shape and growth of the shell in terms of these constant local parameters. These local parameters are important since they, or their information equivalent, is encoded the the DNA of the organism, upon whose instructions the shell is built. The present model is a physical model rather than a descriptive model based on global shell measurements. It has all the attributes of a physical model: It is invariant with respect to position, time, orientation, and choice of coordinate system.

A rather complete review of the history and literature of seashell modeling is given in references \cite{Fowler} and \cite{VanOsselaer}. The global model of Cortie\cite{Cortie} is of special interest. It essentially coincides with the results of the present model. Cortie gives global parameters for many species of mollusk and using the present model, these parameters may be used to derive the local parameters for the listed species.

In this paper, symbols in regular font (e.g. $s$, $\alpha$, $\theta$) will stand for physical scalars, symbols in bold font (e.g. $\vect{r}$, $\vect{v}$) will stand for physical vectors (1st rank tensors) and large bold symbols  (e.g. $\Tens{R}$, $\Tens{S}$, $\Tens{U}$ ) will stand for physical vector operators (2nd rank tensors), unless otherwise noted. These tensors will have different representations in different coordinate systems, with vectors (1st-rank tensors) being generally represented by a 1x3 matrix (e.g. $(u_{x},u_{y},u_{z}))$ and vector operators  (2nd-rank tensors) being represented by a 3x3 matrix. When dealing with only one coordinate system, no distinction will be made between the tensor and its representation, and when dealing with a physical process, the description will be independent of coordinate system. When multiple coordinate systems are being dealt with, a symbolic distinction will be made between the physical tensors and their representations in different coordinate systems.

\section{The Logarithmic Spiral and Helico-spiral}

\noindent A logarithmic spiral is a type of 2-dimension spiral. For a spiral centered at the origin, it may be expressed in 2-dimensional polar coordinates (r,$\theta$) as:

    $r[\theta] =  r[0]\,e^{\cot[\alpha] \theta}$

\noindent where $r[0]$ is the radial distance of the spiral from the origin at $\theta$=0, and the angle $\alpha$ is the angle between the forward-directed tangent to the spiral and the radius vector (from the origin to the spiral) which, for a logarithmic spiral, is constant, constrained by $0 \leq \alpha \leq \pi/2$ \footnote{Alternatively\cite{Illert1989}, negative values of the angle $\alpha$ may be included which will yield a left-handed spiral}. In coordinate-free notation, the radius vector can be expressed as:

\begin{wrapfigure}[16]{l}{0.47\textwidth}
\begin{center}
\includegraphics[width=0.5\textwidth]{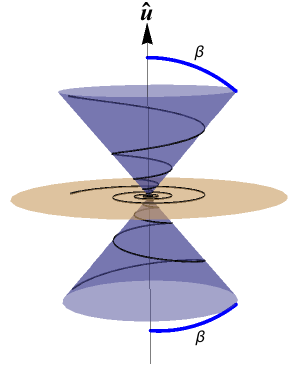}
\end{center}
\caption{Right-handed logarithmic helico-spiral with the projected spiral on the plane of $\nvec{u}$}
\end{wrapfigure}

$\vect{r}[\theta]=e^{\cot[\alpha] \theta}\, \Tens{R}[\theta] \cdot \vect{r}[0]$

\noindent where  $\vect{r}[0]$ is the radius vector at $\theta$=0 and $ \Tens{R}$[$\theta$] is a rotation operator. In a Cartesian coordinate system, where $\vect{r}[\theta]$ is represented by $(x[\theta], y[\theta])$, the rotation operator is represented by:

    $\Tens{R}[\theta] = $ \begin{math} \left(
    \begin{array}{cc}
    \cos[\theta] & -\sin[\theta]\\
    \sin[\theta]	& \cos[\theta]
    \end{array} \right) \end{math}

The rotation operator may be expressed as the exponential of the product of the angle $\theta$ and the generator of the rotation:

     $\Tens{R}[\theta] = e^{\,\tens{U}\theta}$
     
\noindent where the generator is represented as:

     $\Tens{U}=\left( \begin{array}{cc}
    0 & -1\\
    1 & 0
    \end{array}\right)$ 

\noindent This allows the centered logarithmic spiral to be expressed as:

\begin{equation}\label{general}
\vect{r}[\theta]= e^{\tens{S}\,\theta} \cdot  \vect{r}[0]
\end{equation}

\noindent where, defining $\Tens{1}$ as the 2x2 identity operator, $\Tens{S}$ is the operator:

    $\Tens{S} = \cot[\alpha] \Tens{1} + \Tens{U}$
    
The spiral center is at $\theta = -\infty$, (i.e. $\vect{r}[-\infty]=0$). The spiral generated by the above equation is a "right handed" or counter-clockwise spiral. Changing the sign of the $\cot[\alpha]$ term, or the $\Tens{U}$ operator, but not both, will yield a left-handed spiral.

\noindent Equation \ref{general} above may be extended to three dimensions to specify a logarithmic helico-spiral. In three dimensions, the generator of a right-handed rotation of angle $\theta$ about an axis given by a unit vector $\nvec{u} $ represented by $(u_{x},u_{y},u_{z})$ can be represented by:

$\Tens{U} = $ \begin{math} \left(
    \begin{array}{ccc}
    0 & -u_{z} & u_{y}\\
    u_{z} & 0 & -u_{x}\\
    -u_{y} & u_{x} & 0
    \end{array} \right) \end{math}

\noindent and this, along with the 3x3 identity matrix, may be substituted into the expression for $\Tens{S}$ in Eq. \ref{general}, yielding the Cartesian representation of a centered logarithmic helico-spiral. Rather than winding in a plane, the logarithmic helico-spiral winds right-handedly along a double cone with $\nvec{u}$  as the axis of both the cone and the rotation (see Figure 1). The helico-spiral can be specified by it's projection onto the plane perpendicular to $\nvec{u}$, and the cone around which it winds. While the $\alpha$ parameter describes the projected plane spiral, a new parameter $\beta$ is required to specify the cone, where $\beta$ is the angle between the $\nvec{u}$ axis and any vector on the upper cone, or, equivalently, the angle between the -$\nvec{u}$ axis and any vector on the lower cone. This restricts $\beta$ to $0 \leq \beta \leq \pi/2$. The point on the cone at \vect{r}[0] is as good as any other, so $\cos[\beta]$ may be written as:

\begin{equation}\label{betadef}
\cos[\beta] =|\vect{r}[0] \cdot \nvec{u}|/| \vect{r}_{c}[0] |
\end{equation}

\noindent In a more general formulation, the helico-spiral may be displaced and rotated from its "standard" position described above, without altering its shape.

\section{The Differential Equation of Shell Growth}

One of the purposes of this model is to describe the shell from the point of view of a hypothetical organism building it. The edge of the shell aperture forms a closed, oriented curve in space and, for the purposes of this model, the organism is assumed to live on the aperture and, for building purposes, is only "aware" of the aperture as it exists at the present time. Its building "instructions" consist of a set of internalized constant parameters which determine, given the present aperture, the varying rate and direction of growth at each point on the aperture edge. These parameters, or their information equivalent, are mirrored in the organism's DNA. These parameters are "local" to the organism, as opposed to "global" parameters which are used describe the shell in terms of measurements made by a researcher on the entire shell.

For building purposes, the organism doesn't know its history or its orientation in space. It lives on the aperture and knows the aperture, and it knows what the aperture should be an infinitesimal time $dt$ later and it builds new shell accordingly. It is assumed that the organism simply grows shell at its aperture edge at various predetermined rates (or possibly relative rates: slower growth rates when times are bad), without reference to any global considerations, such as the condition or shape of the entire shell. Using the present aperture as a measuring stick, it always "sees" the same thing and this is the essence of the self-similarity of the shell structure.

Mathematically, it is assumed that the aperture is defined in terms of a right-handed Cartesian coordinate system ($x$,$y$,$z$) with the $xy$ plane being the "aperture plane" and the $z$ axis being normal to that plane, pointed roughly in the direction of growth, "out" of the shell. The aperture edge is described by the aperture function, also known as the generating curve. The aperture function is a vector function $\vect{f}\,[s] = (f_{x}[s], f_{y}[s], f_{z}[s])$,  parametric in a variable $s$, and describes the aperture edge at time  $t=0$. For example, if the aperture edge were a circle of radius $\rho$, it could be described in the aperture coordinate system by $\vect{f}\,[s]=\rho\,(\cos[s],\sin[s],0)$ where $s$, in this case, is the angle from the local $x$-axis to a point on the circle. $\vect{f}\,[s]$  will also contain any number of  parameters, such as $\rho$ in the above example, to describe the standard (time zero) aperture edge. It is assumed that the aperture at an infinitesimal time later is described by an infinitesimal advance of the aperture, magnification of the aperture, and rotation of the aperture. Mathematically, the aperture at time $t+\Delta t$ will be described by:

    $\vect{r}[t+\Delta t] = e^{\sigma \Delta t}\, \tens{R}[\nvec{u},\omega \Delta t] \cdot \vect{r}[t] + \vect{A}[\Delta t]$

\noindent where $e^{\sigma \Delta t}$ is the exponential enlargement of the aperture in time $\Delta t$, $\Tens{R}[$\nvec{u}$,\omega \Delta  t]\cdot \vect{r}[t]$ is the rotation of $\vect{r}[t] $ in time $\Delta t$, with $\omega$ being the rotation rate, and $\vect{A}[\Delta t]$ is the vector displacement of the center of the aperture in time $\Delta t$,

In the limit of infinitesimal $\Delta t$, the aperture enlargement may be approximated to first order as:

    $e^{\sigma  \Delta t } \rightarrow  1 + \sigma  dt$
    
The rotation can be described as a rotation by angle $\theta=\omega t$ about an axis specified by a unit vector $\nvec{u}=(u_{x},u_{y},u_{z})$. Using the Rodrigues rotation formula:


    $\Tens{R}[\nvec{u},\theta]\!=\!\Tens{1}+\sin[\theta] \Tens{U}+(1-\cos[\theta])\Tens{U}\cdot\Tens{U}$
      
\noindent where $\Tens{1}$ is the 3x3 identity matrix and $\Tens{U}$ is the generator of the rotation, derived from $\nvec{u}$:

    $\Tens{U} = $ \begin{math} \left(
    \begin{array}{ccc}
    0 & -u_{z} & u_{y}\\
    u_{z} & 0 & -u_{x}\\
    -u_{y} & u_{x} & 0
    \end{array} \right) \end{math}
    
\noindent Infinitesimally, such a rotation may be represented as: 

     $\Tens{R}[\nvec{u}, \omega\Delta t] =  e^{\tens{U}\,\omega \Delta t} \rightarrow \Tens{1}+ \omega \Tens{U} dt$.
     
\noindent The displacement vector $\vect{A}[\Delta t]$ can be described infinitesimally as:

    $L[\Delta t]  \rightarrow  \vect{v}\, dt$
    
\noindent where $\vect{v}$ is the forward displacement velocity of the aperture center, assumed to be constant from the aperture point of view. The infinitesimal growth equation may now be written as:

      $\vect{r}[t+dt] =(1+\sigma  dt)( \Tens{1}+\omega \Tens{U}\,dt) \cdot \vect{r}[t] + \vect{v} dt$
      
\noindent which yields the first order differential equation for the shell:

    \begin{equation}\label{diffeq}
    d\vect{r}[t]/dt= (\sigma \Tens{1}+\omega \Tens{U})  \cdot  \vect{r}[t] + \vect{v} 
    \end{equation}

\noindent This is the equation which describes the growth of the shell in terms of parameters $\sigma$, $\omega$, $\nvec{u}$ and $\vect{v}$ which, from the point of view of the organism, moving with the aperture, are constant.

Define the operator $\Tens{S}=(\cot[\alpha] \Tens{1}+ \Tens{U})$ and its inverse $\Tens{T}$, where $\Tens{S} \cdot \Tens{T}=\Tens{1}$ and where $\cot[\alpha]=\sigma/\omega$. The solution to the above equation is then of the form:

    $r[t] = e^{ \tens{S}\,\omega t}\cdot \vect{K} - \Tens{T} \cdot \vect{v}$
    
\noindent where $\vect{K}$ is a vector that is constant in time, and is determined by boundary conditions. Since the vector function \vect{r}[0]=\vect{f}\,[s] is the  equation of the aperture edge at $t=0$,  the solution is:

\begin{equation}\label{inteq}
\vect{r}[t,s] =  e^{\tens{S}\,\omega t} \cdot  (\Tens{T} \cdot \vect{v}+ \vect{f}\,[s] )  - \Tens{T} \cdot \vect{v} 
\end{equation}

This is the coordinate-free equation of the shell with general orientation and displacement. Making the substitutions $\omega t  \rightarrow  \theta$, $\sigma /\omega  \rightarrow  \cot[\alpha]$, $\vect{f}\,[s]  \rightarrow  \vect{r}[0,s]$ it can be seen that  $\vect{r}[t,s]$ describes a family of logarithmic helico-spirals, parametric in $s$, all  originating at $-\Tens{T} \cdot \vect{v}$ growing along the $\pm \nvec{u}$  axis. When the sign is positive, the shell grows on the upper cone with a right-handed twist, and when it is negative, the shell grows on the lower cone with a left handed twist. These spirals specify the two-dimensional surface of the shell. The origin of all spirals is the same:

    $\vect{r}[-\infty,s]  = -\Tens{T}\cdot \vect{v}$

\noindent The shell centerline is given by setting $\vect{f}\,[s] \rightarrow 0$:

    $\vect{r}_{c}[t] =  e^{\tens{S}\,\omega t} \cdot  \Tens{T} \cdot \vect{v} -\Tens{T} \cdot \vect{v}$

\noindent In the following discussion, the $\omega t$ term will be replaced by the angle $\theta$, which is seen to be the angle of the projection of the centerline onto the plane of $\nvec{u}$. The angle $\theta$ will still be referred to as the "time".

\section{Local and Global Coordinate Systems}

The solution $\vect{r}[\theta,s]$ in the present model (Equation \ref{inteq}) is in coordinate-free notation. In order to actually perform calculations,  a set of coordinate systems must be chosen in order to represent the various vectors and tensors as matrices. A global coordinate system is one in which the shell is motionless and in which measurements of the shell may be made. A local coordinate system moves with the aperture center, in which the aperture center is unvarying in time, and in which the aperture function undergoes only magnification in time, and in which organism lives. In the local system, representations of of various parameters such as the axis of rotation and the velocity take on constant values and constitute the unvarying "instructions" by which the organism builds the shell.  An aperture system is defined as a local system derived from the aperture itself and the organism's orientation in that aperture, and is the only system available to the organism. Other local systems, derived from various vector quantities which are constant in the local system may also be defined (See e.g Appendix I), but are not accessible to the organism.

First, we may choose an aperture coordinate system, which is local,  based on the aperture in some convenient way. For example, for an elliptical aperture, we may say that the ellipse lies on the aperture $xy$ plane with $\vect{f}[0]$ specifying the $x$ axis, and the $z$ axis is normal to that $xy$ plane, forming a right-handed coordinate system. The $z$-axis will be such that the $z$ component of the forward growth vector $\vect{v}$ is positive. If $\vect{f}\,[s]$ is set to zero, $\vect{r}[t]$ will specify the origin of that aperture coordinate system. The $\nvec{u}$ and $\vect{v}$ vectors will have a concrete and constant representation in this system, and the solution for $\vect{r}[t]$ will yield the global shell in the aperture system. These technical concerns are of no concern to the organism, but rather are the result of our mathematical description of how the organism builds its shell.

We may next choose a global coordinate system in which the shell as a whole is at rest.  It is convenient to choose the global origin to be the spiral origin at $\vect{r}[-\infty]$. Although it is not necessary, it is convenient to choose the global $z$-axis to be parallel with the $\nvec{u}$ vector, which is constant in the global system as well, since it is an eigenvector of the rotation portion of the $\Tens{S}$ operator. The representation of $\nvec{u}$ will differ depending upon global or local coordinate system, but the vector itself is constant in time in either system. We may choose a right-handed Cartesian coordinate system and set the origin of the aperture ($\vect{r}_{c}[0]$) at $\theta=0$ to lie in the global $xz$ plane, such that the $x$ component of the aperture origin is positive. The $z$ component of $\vect{r}_{c}[0]$ for right- and left-handed shells will be positive and negative respectively. In the global system,  the $\nvec{u}$ vector will be unmoving, but the $\vect{v}$ vector will be moving, always equal to the velocity of the centerline $(\vect{v}=d\vect{r}_{c}[t]/dt)$ in the global system. In a local system, both representations will be constant in time.

\noindent In the following, The $P$ subscript will stand for matrix representations in the particular global coordinate system chosen by the present model.  The $A$ subscript will stand for representations in the particular local coordinate system chosen by the present model, which is an aperture coordinate system. In the spiral-centered global system of the present model, the equation the shell is:

    $\vect{r}_{P}[\theta,s] =  e^{ \tens{S}_{P} \theta}\cdot  (\Tens{T}_{P}\cdot \vect{v}_{P}+ \vect{f}_{P} [s] ) $
  
\noindent while in the aperture system it is:
  
     $\vect{r}_{A}[\theta,s] =  e^{ \tens{S}_{A} \theta} \cdot  ((\Tens{T}_{A}\cdot \vect{v}_{A}+ \vect{f}_{A}[s] )  - \Tens{T}_{A}\cdot \vect{v}_{A}$

In the present model, the $\nvec{u}$ axis is chosen to coincide with the global $\nvec{z}$ axis which yields $\vect{u}_{P}=(0,0,1)$. $\Tens{S}_{P}$  is then equal to $\cot[\alpha] \Tens{1} +  \Tens{U}_{P}$, where $ \Tens{U}_{P}$, is derived from $\vect{u}_{P}$ in the usual way, and  $\Tens{T}_{P}$ is then the matrix inverse of $\Tens{S}_{P}$. The centerline at time zero is given by $\vect{r}_{cP}[0]=\Tens{T}_{P}\cdot \vect{v}_{P}$ which is set equal to $A\,(\sin[\beta], 0, D \cos[\beta])$, and $\vect{f}_{P}[s]$ is the parametric representation of the aperture function in the present global coordinate system. In aperture coordinates, $\vect{u}_{A}=(u_{x},u_{y},u_{z})$ allowing $\Tens{U}_{A}$, $\Tens{S}_{A}$, and $\Tens{T}_{A}$ to be calculated, with $\vect{f}_{A}[s]$ being the parametric representation of the aperture function in aperture coordinates, which is usually the simplest description of the aperture function.  The constant local growth vector is given by $\vect{v}_{A} = (v_{x},v_{y},v_{z})$.

Except for position vectors, the local and global vector and operator representations will in general be related by some rotation $\Tens{R}_{o}$ such that for any vector $\vect{q}$,  $\vect{q}_{P} = \Tens{R}_{o} \cdot \vect{q}_{A}$.  It is desirable to express the aperture function in local terms, where it takes its simplest mathematical form. The global equation then will be:

    \begin{equation}\label{Ro}
    \vect{r}_{P}[\theta,s] =  e^ {\tens{S}_{P} \theta}\cdot  (\Tens{T}_{P} \cdot \vect{v}_{P}+ \Tens{R}_{o} \cdot \vect{f}_{A}[s] )
    \end{equation}

\section{Local and Global Parameters}

The global parameters are a set of six easily-measured parameters which describe the shell centerline. The global parameters are the angles $\alpha$ and $\beta$ described above, $A$,  $D$, and the components of the axis of rotation $\nvec{u}_{P}$. $A$ is the distance from the cone origin to the centerline position at time zero $\vect{r}_{c}[0]$, and $D$ is equal to 1 for a right-handed spiral and -1 for a left-handed spiral. The representation of the unit vector $\nvec{u}$, the axis of rotation, must be known as well, which requires two more constants.

The local parameters are a set of six parameters which the organism uses to define its centerline development. The local parameters are $\alpha$,  $\nvec{u}$ and  $\vect{v}$. $\alpha$ is counted as both  a global and local parameter, while the centerline velocity $\vect{v}$, consisting of three independent parameters, is purely local. The representation of the unit vector $\nvec{u}$, the axis of rotation, must be known as well, which requires two more constants.

\subsection{The Global $\beta$ Parameter}

Assuming $0<\alpha<\pi/2$ and $0<\beta<=\pi/2$, define the local parameter $\gamma$:

    $\cos[\gamma] = D\, \nvec{u}.\vect{v} / |\vect{v}|  = |\nvec{u}.\vect{v}|/|\vect{v}|$
    
\noindent which is the angle between $\vect{v}$ and the growth axis $D \nvec{u}$. The angle $\gamma$ is restricted by $0 \leq \gamma \leq \pi/2$. The angle $\beta$ is the angle between the $\nvec{u}$ axis and a point on the spiral centerline for the upper cone, and between the $-\nvec{u}$ axis and the centerline on the lower cone. $\beta$ is thus restricted to $0\leq\beta\leq\pi/2$. From Equation \ref{betadef}, the time-zero centerline $\vect{r}_{c}[0]$ can be used to define $\beta$. In local parameter terms:

    $\cos[\beta]=|\vect{r}_{c}[0].D\nvec{u}|/ |\vect{r}_{c}[0]| = |(\Tens{T}.\vect{v}) .D\nvec{u}|/|\Tens{T}.\vect{v}| = \cos[\gamma]\sqrt{\cos[\alpha]^2 + \sin[\alpha]^2\cos[\gamma]^2}$

\noindent Since $0 \leq \beta \leq \pi/2$ and $0 \leq \gamma \leq \pi/2$, the other trigonometric functions are easily calculated and $\beta$ may be calculated from local parameters by:

    $\beta=\arctan[\cos[\alpha] \tan[\gamma]]$

\subsection{The Global $A$ Parameter}

The $A$ parameter is the magnitude of the vector from the spiral origin to the centerline at time zero.

   $A = |\vect{r}_{c}[0]| = |\Tens{T}.\vect{v}|  = |\vect{v}|\sin[\alpha]\sqrt{1+\cos[\gamma]^2 \tan[\alpha]^2}$

\subsection{The Global $D$ Parameter and Direction of Growth}

The double cone on which the shell winds presents two growth possibilities: Growth upward from the origin at $\theta=-\infty$, with a right-handed rotation along the positive $\nvec{u}$ axis, or growth downward with a left-handed rotation along the $-\nvec{u}$ axis. In other words, the growth axis may be either the $+\nvec{u}$ or the $-\nvec{u}$ axis\footnote{In Cortie's model below, a parameter $D=\pm 1$ is specified \textit{a priori} to handle these two cases. A negative value of $D$ effectively inverts the global coordinate system, so that all growth is downward, right-handedly for $D=1$ and left-handedly for $D=-1$.}. In the present model, both growth directions are accepted, and the $D$ parameter is derivable from the local parameters. If the $\vect{v}$ vector points above the plane of $\nvec{u}$, growth will be upward and right-handed, while if it points below, growth will be downward and left-handed. If $\vect{v}$ lies in the plane of $\nvec{u}$, there will be no growth along an axis, only outward. Mathematically, the $D$ parameter may be expressed as:

    $D = \nvec{u}\cdot \vect{v} / |\nvec{u}\cdot \vect{v}|$ 

The present model, when expressed terms of $\nvec{u}$ and $\vect{v}$ may use the $D$ parameter, with the understanding that it is not fundamental, but is rather derived from the fundamental $\nvec{u}$ and $\vect{v}$ vectors. This viewpoint of $D$ is necessary for the coordinate-free representation of the shell growth which is in turn necessary to express the growth process as a physical process, which in fact, it is.
    
\subsection{The Global and Local $\vect{v}$ Parameter}

The centerline velocity at time zero is given by:

    $\vect{v}=\Tens{S}\cdot\vect{r}_{c}[0]$
    
\noindent $\Tens{S}$ is already defined using global parameters, so the $\vect{r}_{c}[0]$ vector must be expressed in global parameters. Since $\vect{r}_{c}[0]$ lies on the cone, for a coordinate system centered on the cone origin, $\vect{r}_{c}[0]$ may be represented as:

    \begin{equation}\label{rco}
    \vect{r}_{c}[0] = A (\sin[\beta] \nvec{\rho}+ D \cos[\beta] \nvec{u})
    \end{equation}

\noindent where $D \cos[\beta] \nvec{u}$ is the component of $\vect{r}_{c}[0]$ along the $\nvec{u}$ axis and $\nvec{\rho}$ is some unit vector in the plane of $\nvec{u}$. 

\section{The Present Model}

The present model in global terms chooses the central axis of the shell to be the positive $\nvec{z}$ axis so that $\nvec{u}_{P}=(0,0,1)$, and specifies the location of the centerline at time zero to lie in the $xz$ plane. Thus, using Equation \ref{Ro},

    \begin{equation}\label{PresentModel}
    \vect{r}_{P}[\theta, s] = e^{\tens{S}_{P}\,\theta} \cdot  (\Tens{r}_{oP}  + \Tens{R}_{o} \cdot  \vect{f}_{A}[s] )
    \end{equation}

\noindent where
    
    $\Tens{U}_{P} = $ \begin{math} \left(
    \begin{array}{ccc}
    0 & -1 & 0\\
    1 & 0 & 0\\
   0 & 0& 0
    \end{array} \right) \end{math}

\noindent  and

    $\Tens{S}_{P}= \cot[\alpha]  \Tens{1} +  \Tens{U}_{P}$

\noindent places a right handed spiral on the upper cone and a left handed spiral on the lower cone. From Equation \ref{rco}, the aperture center at time zero is:

\begin{wrapfigure}[23]{l}{0.5\textwidth}
\begin{center}
\includegraphics[width=0.5\textwidth]{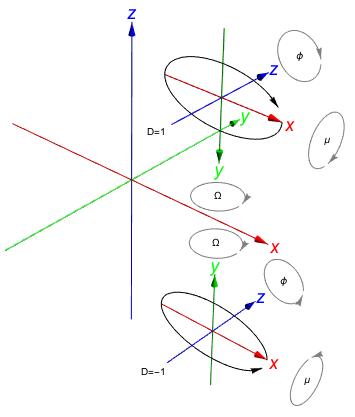}
\end{center}
\caption{Elliptical aperture functions for the present model at time zero. The rotation directions are for increasing $s$. Also shown are the directions of rotation induced by $R_{o}[\mu,\phi,\Omega]$  for each of the angles $\mu$, $\phi$, and $\Omega$. Note that the $xy$ plane divides the $D=1$ and $D=-1$ cases into mirror images of each other, reversing the local $y$-axis.}
\end{wrapfigure}

    $\Tens{r}_{cP}[0] = \Tens{T}_{P} \cdot \vect{v}_{P} = A\, (\sin[\beta],0,D\,\cos[\beta])$

\noindent and  $\Tens{R}_{o}$ is an as yet undetermined transformation matrix which converts the local aperture function $\vect{f}_{A}[s]$ to the global aperture function $\vect{f}_{P}[s]$.  Generally, $\vect{f}_{A}[s]$ may be chosen to be any cyclic continuous vector function of $s$ for which $\vect{f}\,[0]=\vect{f}\,[2\pi]$. As an example, if the aperture is elliptical in shape, then

    $\vect{f}_{A}[s] = R[s]\,\nvec{n}[s]$

\noindent where $a$ and $b$ are constants and:

    $R[s]=1/\sqrt{\frac{\cos[s]^2}{a^2} + \frac{\sin[s]^2}{b^2}}$
    
    $\nvec{n}[s]=(\cos[s],\sin[s],0)$

The transformation matrix $\Tens{R}_{o}$ converts any local tensor representation to its global representation, including $\nvec{u}_{A}$ and $\vect{v}_{A}$:

    $\Tens{R}_{o}\cdot\nvec{u}_{A}=\nvec{u}_{P}$
    
     $\Tens{R}_{o}\cdot\vect{v}_{A}=\vect{v}_{P}$

\noindent which is sufficient to define $\Tens{R}_{o}$ in terms of $\nvec{u}_{P}$, $\nvec{v}_{P}$, $\nvec{u}_{A}$, and $\nvec{v}_{A}$, but the representation is rather cumbersome and will not be written here, although an algorithmic representation of $\Tens{R}_{o}$ is easily implemented (See Appendix II). As shown in the next section, Cortie's model can be used to express $\Tens{R}_{o}$ numerically using his provided measurements on actual shells.

\section{Cortie's Model}

The model developed by Cortie\cite{Cortie} describes the shell in a global coordinate system using seven fundamental parameters, plus any number of parameters needed to describe the aperture function (e.g. the major and minor axes of an elliptical aperture). The seven main parameters are the four global parameters $\alpha$, $\beta$, $D$ and $A$, along with three angles,  $\phi$, $\mu$, and $\Omega$. As in the present model, his model is a function of $s$ and $\theta$. Cortie's equations for the shell had some typographical errors, which were corrected by Picado\cite{Picado}, who also provides a \textit{Mathematica} program for calculating Cortie's model. Cortie's corrected equations are:

    \begin{equation}
     \begin{array}{l}
    x[\theta, s] = D\,e^{\cot[\alpha] \theta}\, ( A \sin[\beta] \cos[\theta] + \\ 
    R \cos[s\!+\!\phi] \cos[\theta\!+\!\Omega] - R \sin[\mu] \sin[s\!+\!\phi] \sin[\theta\!+\!\Omega] )
     \end{array}
    \end{equation}

    \begin{equation}
    \begin{array}{l}
    y[\theta, s] =e^{\cot[\alpha] \theta}\,( -A \sin[\beta] \sin[\theta] - \\
    R \cos[s\!+\!\phi] \sin[\theta\!+\!\Omega] - R \sin[\mu] \sin[s\!+\!\phi] \cos[\theta\!+\!\Omega] )
    \end{array}
    \end{equation}

    \begin{equation}
     \begin{array}{l}
    z[\theta, s] = e^{\cot[\alpha] \theta} \, ( -A \cos[\beta]  + R  \sin[s\!+\!\phi]  \cos[\mu] )
     \end{array}
    \end{equation}

Cortie's model is expressed in terms of a global coordinate system and yields essentially the same numerical results as the present model. Cortie's model can be expressed in a concise form in a way very similar to the present model. Cortie's $R$ function is better written $R[\theta,s]$ and is a function of $\theta$ only to implement periodic knobs on the shell which will not be considered in this model, so the dependence of $R[\theta,s]$ upon $\theta$ will be removed. Also, $R[s]$ should properly be considered to involve any number of additional parameters needed to describe the aperture function. Defining $\vect{r}_{C}[\theta,s] = (x[\theta, s], y[\theta, s], z[\theta, s])$,  and using the subscript $C$ to indicate representations in Cortie's model, his model may be expressed as:

\begin{equation}\label{Cortie1}
\vect{r}_{C}[\theta, s] = \Tens{D} \cdot e^{\tens{S}_{C} \theta} \cdot  (\, \vect{r}_{c\,C}[0]  + \Tens{R}_{oC}[\mu,\phi,\Omega] \cdot \vect{f}_{LC}[s]\,)
\end{equation}
    
\noindent where $\Tens{S}_{C} = \cot[\alpha] \Tens{1} +\Tens{U}_{C}$ , $\Tens{U}_{C}$ being the generator of a rotation associated with the $-z$ axis vector  $\nvec{u}_{C}=(0,0,-1)$:

    $\Tens{U}_{C} = $ \begin{math} \left(
    \begin{array}{ccc}
    0 & 1 & 0\\
    -1 & 0 & 0\\
   0 & 0& 0
    \end{array} \right) \end{math}
    
and $\vect{f}_{LC}[s]$ represents Cortie's expression for the aperture function in local terms.

The $ \Tens{D}$ operator inverts the $x$ axis for $D = -1$. Cortie inverts the $x$ axis  when D=-1 to generate a left-handed coordinate system, so that both left- and right-handed cases can be implemented on either the lower or the upper cone. Cortie chooses the lower cone since this is standard in the literature, but the  $ \Tens{D}$  operator is not physical, which means the above equation is not a physical equation.

\noindent The  $ \Tens{D}$  operator is given  by:

   $\Tens{D} = $ \begin{math} \left(
    \begin{array}{ccc}
   D&0& 0\\
   0 & 1 & 0\\
   0 & 0& 1
    \end{array} \right) \end{math}
    
\noindent $\Tens{D}\cdot \vect{r}_{c\,C}[0]$ is the location of the aperture origin at time zero always lying on the lower cone:

    $\vect{r}_{c\,C}[0] = A ( \sin[\beta], 0, -\cos[\beta] )$

\noindent Cortie's global aperture function is parametric in $s$ and is given by  $\Tens{R}_{oC}[\mu,\phi,\Omega] \cdot (R[s] \nvec{n}[s])$, where $\Tens{R}_{oC}[\mu,\phi,\Omega]$  is a rotation specified by angles  $\mu$,  $\phi$, and $\Omega$. It can be expressed as the product of three rotations\cite{Davenport}:

$\Tens{R}_{oC}[\mu,\phi,\Omega]$= \begin{math} 
\left( \begin{array}{ccc}
\cos[\Omega]&\sin[\Omega]&0\\
-\sin[\Omega]&\cos[\Omega]&0\\
0&0&1
\end{array}\right)
\cdot
\left(\begin{array}{ccc}
1&	0&	0\\
0&	\cos[\mu]	&-\sin[\mu]\\
0	&\sin[\mu]	&\cos[\mu]
\end{array}\right)
\cdot
\left(\begin{array}{ccc}
\cos[\phi]&	0	&-\sin[\phi]\\
0	&1	&0\\
\sin[\phi]&	0&	\cos[\phi]
\end{array}\right)
\end{math}
     
\noindent where the three angles are generally restricted by $-\pi\! < \!\phi\! \leq \!\pi$ , $-\pi \!< \! \mu \leq \pi $, and $-\pi/2 < \Omega \leq  \pi/2$.

It can be seen that the angles $\mu$, $\phi$ and $\Omega$ correspond to rotations around the $x$, $y$ and $z$ axis respectively in a manner similar to a Tait-Bryan expression for a 3-D rotation\cite{Davenport}, with $\Tens{R}_{oC}[0,0,0]=1$. $R[s] \nvec{n}[s]$ is Cortie's expression for the local aperture function

    $\vect{f}_{LC}=R[s] \nvec{n}_{C}[s]$

\noindent where 

   $\nvec{n}_{C}[s] =(\cos[s], 0, \sin[s])$

\noindent This expresses the aperture in terms of an aperture plane on the $xz$ plane with the normal in the $y$ direction. Cortie's model could be extended to specify $\vect{f}_{LC}$ as a general vector function of parameter $s$, which need not be an angle in the $xz$ plane. This is the case with the present model, in which the aperture function is described in terms    of an aperture

\begin{wrapfigure}[27]{l}{0.5\textwidth}
\begin{center}
\includegraphics[width=0.5\textwidth]{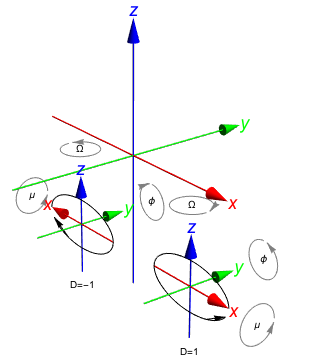}
\end{center}
\caption{Elliptical aperture functions for Cortie's model at time zero. The rotation directions are for increasing $s$. Also shown are the directions of rotation induced by $R_{oC}[\mu,\phi,\Omega]$ for each of the angles $\mu$, $\phi$, and $\Omega$. Note that the $yz$ plane divides the $D=1$ and $D=-1$ cases into mirror images of each other, reversing the local $x$-axis.}
\end{wrapfigure}

\noindent plane on the $xy$ plane with the normal in the $z$ direction, in local coordinates, and is not restricted to the aperture plane.

While Cortie's model expresses $\vect{r}_{C}[\theta,s]$ using the above mentioned seven global parameters, the present model uses four local parameters ($\alpha$ and $\vect{v}$) and two chosen parameters $\nvec{u}$. Since $\nvec{u}$ is normalized, it has only two independent parameters, and the angle $\alpha$ is counted as both a global and local parameter.

Also, Cortie inverts the global coordinate system using the extra $D$ parameter in order to express both right and left-handed rotations on the lower cone of the helico-spiral. While the model is accurate, it is designed to describe the shell using easily-measured global parameters, rather than the parameters that the organism is using to build the shell. Also, the treatment of the handedness of rotation is rather artificial. The actual physical description of the shell must be coordinate-free, and must not depend upon inverting a coordinate system to describe different cases.


Cortie's model can be related to the present model (see Equation \ref{PresentModel}):

    \begin{equation}\label{Cortie2}
    \vect{r}_{C}[\theta, s] = \Tens{J}\cdot e^{\tens{S}_{P}\theta} \cdot  (\vect{r}_{cP}[0]  + \Tens{R}_{o} \cdot  \vect{f}_{A}[s] ) =  \Tens{J}\cdot  \vect{r}_{P}[\theta, s] 
    \end{equation}
   
\noindent  This is derived using:

    $\vect{r}_{c\,C}[0]=\Tens{g}_{1} \cdot \vect{r}_{cP}[0]$
    
    $\vect{f}_{LC}[s]=\Tens{g}_{2} \cdot \vect{f}_A[s]$
    
\noindent where

    $\Tens{g}_{1} = $ \begin{math} \left(
    \begin{array}{ccc}
    1 & 0 & 0\\
    0 & 1 & 0\\
    0 & 0 & -D
    \end{array} \right) \end{math}

   $\Tens{g}_{2} = $ \begin{math} \left(
    \begin{array}{ccc}
    1 & 0 & 0\\
    0 & 0 & 1\\
    0 & 1 &0
    \end{array} \right) \end{math}
    
\noindent It can be shown that:

    $\vect{r}_{C} = \Tens{J} \cdot e^{\tens{S}_{P} \theta}(\vect{r}_{cP}[0]+\Tens{g}_{1} \cdot \Tens{R}_{oC} \cdot \Tens{g}_{2}^{-1} \cdot \vect{f}_{A})$

\noindent where

    $\Tens{J}=\Tens{D} \cdot \Tens{g}_{1} \cdot e^{-2 \tens{U}\, \theta}$

\noindent so that the transformation matrix of the present model $\Tens{R}_{o}$ of equation \ref{PresentModel} is related to Cortie's transformation matrix $\Tens{R}_{oC}$ by:

    $\Tens{R}_{o}=\Tens{g}_{1} \cdot \Tens{R}_{oC} \cdot \Tens{g}_{2}^{-1}$

\section{Calculating Local Parameters from Global Parameters}

In the present model, $\Tens{R}_{o}[\mu,\phi,\Omega]$ transforms the $\vect{f}_{P}[s]$ vector representation to $\vect{f}_{A}[s]$, and is thus the coordinate transformation of any local vector representation to the global representation. Specifically,

\begin{equation}\label{G2L1}
\nvec{u}_{P}[s]=\Tens{R}_{o}[\mu,\phi,\Omega]\cdot \nvec{u}_{A}
\end{equation}

\begin{equation}
\vect{r}_{cP}[0]=\Tens{R}_{o}[\mu,\phi,\Omega]\cdot \vect{r}_{cA}[0]
\end{equation}
   
\noindent Since $\vect{r}_{cP}[0] = \Tens{T}_{P}\cdot vP$, and $\vect{r}_{cP} = A\,{\sin[\beta], 0, D \cos[\beta]}$ is known, it follows that $\vect{v}_{P}$ can be determined by $\vect{v}_{P}=\Tens{S}_{P}\cdot \vect{r}_{c}[0]$ and

\begin{equation}\label{G2L2}
\vect{v}_{P}=\Tens{R}_{o}[\mu,\phi,\Omega]\cdot \vect{v}_{A}
\end{equation}
     
Equations \ref{G2L1} and  \ref{G2L2} above make the connection between the $\phi$, $\mu$, $\Omega$ parameters and the local and global parameters of the present model, rather than leaving them as fundamental parameters as in Cortie's model. Inverting $\Tens{R}_{o}$, the above two equations are sufficient to determine the local parameters used by the organism to build the shell. 

  $\vect{u}_{A}=\Tens{R}_{o}^{-1}.\vect{u}_{P}$
  
  $\vect{v}_{A}=\Tens{R}_{o}^{-1}.\vect{v}_{P}$

\noindent which yields the local parameters $\vect{u}_{A}$ and $\vect{v}_{A}$ in terms of the global, easily measurable, parameters. Conversely, the $\mu$, $\phi$, $\Omega$ angles can, in principle, be expressed in terms of $\vect{u}_{P}$ and$\vect{v}_{P}$, $\vect{u}_{A}$ and $\vect{v}_{A}$, although the symbolic representation for $\Tens{R}_{o}$ in these terms is rather cumbersome.  In the \textit{Mathematica} code in Appendix II, the transformation matrix $\Tens{R}_{o}$ is calculated numerically from input values of $\vect{u}_{P}$, $\vect{u}_{A}$, $\vect{v}_{P}$, and $\vect{v}_{A}$.

\section{The Aperture Function}

The aperture function describes the aperture edge at time  $\theta=0$. It is a vector function $(\vect{f}\,[s]=(f_{x}[s],f_{y}[s],f_{z}[s])$ which describes a closed curve, parameterized by $s$ and it is assumed that $\vect{f}\,[0]=\vect{f}\,[2\pi]$. It is in the aperture system that the aperture function usually assumes its simplest expression. For simple cases (e.g. circle and ellipse), $s$ may be defined as the angle in the aperture plane from the $s=0$ point, which may be taken to be the $x$-axis in the aperture plane, but in more complicated cases (e.g. a "U" shaped aperture) using $s$ as an angle may yield an unusable double-valued function. The $s$ parameter may be chosen by convenience, and it is assumed that it is monotonically increasing with distance along the aperture curve.

For a given aperture function. there can be a certain ambiguity in the definition of the aperture plane. For example, a circular aperture, or any aperture contained in a plane, suggests that plane as the choice of the aperture plane, with $\vect{f}_{z}[s]$ being equal to zero, but the aperture plane could just as easily be defined as being tilted and displaced with respect to the plane of the circle. This would entail $\vect{f}_{z}[s]$ being generally non-zero. For apertures which vary in three dimensions, for which $\vect{f}_{z}[s]$ is unavoidably non-zero, it is not clear how the aperture function defines the aperture plane, and thus the coordinate system for the aperture. This ambiguity results in an ambiguity in the $\vect{v}$ and $\vect{u}$ vectors, in that part of the displacement and rotation specified by the $\vect{u}$ and $\vect{v}$ vectors can be removed and placed into the aperture function. There can be various different choices of coordinate system and aperture function which yield exactly the same global results. In order to compare different shells, a consistent method of determining the aperture coordinate system is needed which makes no reference to global parameters. 

One method to eliminate this ambiguity and to make the $\vect{u}$ and $\vect{v}$ vectors unambiguous is to require that the aperture function have no net displacement or rotation in the aperture frame. Mathematically, this would mean that, in the aperture system,

    $\oint \vect{f}\,[s]\,df= \vect{0} $
    
    $\oint f_{i}[s]\, f_{j}[s]\,df =0$ for $i \neq j$
      
\noindent where $f$ is the distance along the aperture curve from $s=0$ and

    $df = \left| \frac{\partial\,\vect{f}\,[s]}{\partial s}\right|\,ds$

In other words, the description provided by $\vect{f}\,[s]$ is constrained such that the average of the aperture function defines the origin of the aperture coordinate system\footnote{This is essentially the same definition used by Illert\cite{Illert1989} who expressed the constraint as $\vect{r}_{c}[\theta]=\oint \vect{r}\,[\theta,s]\,df= \vect{0}$ in a global system.}, and the plane of minimum variance defines the aperture plane.\footnote{It is assumed that the variation of the aperture function in the $z$ direction is less than the variation in the $xy$ plane.} For example, in the case of an elliptical aperture function, this would place the origin at the center of the ellipse, and the aperture plane would be the plane of the ellipse. 

Another method to remove the ambiguity is to specify the aperture plane by using one of the two constant vectors available to the organism, namely the $\nvec{u}$ or $\vect{v}$ vector. Using the aperture average as the local origin, one of these two vectors could be specified as the aperture z-axis with the plane normal to them as the aperture plane. The problem with this approach is that the organism only has access to the aperture frame, and the information contained in the representations of $\nvec{u}$ and $\vect{v}$ are used to build the shell. A representation that uses one of these vectors will yield a trivial, and therefore useless representation of that vector.

A further issue is the problem of how the organism maintains its orientation in the aperture plane. In mathematical terms, the question is "how does the organism define and maintain the $x$ and $y$ axes in the aperture coordinate system?". The organism must store the information equivalent of the $\nvec{u}$ and $\vect{v}$ vectors, but in order to do so, they must be referenced to a coordinate system which is not defined by them, or any other vector derived from them and the $\alpha$ angle.

In the present model, the $x$-axis is arbitrarily chosen as the s=0 point in a mathematically convenient (and to that extent arbitrary) definition of the aperture function. The organism itself surely has an internal sense of direction in the $xy$ plane but it would have to be quite immune to any disturbance of that sense in order to produce the very accurate shells seen in nature, without some sort of self-correcting mechanism. The aperture shape itself might be used in conjunction with the internal orientation to maintain orientation, but this would be impossible for a circular aperture. It may be that the idea that the organism has no sense of the global shell, and senses only the aperture, is not correct. The organism may  be in contact with part of the shell that was created at a time roughly $\theta=2\pi$ previously, and may use that contact region as a point of orientation. 

With regard to the idea that the organism uses the aperture shape to determine orientation in the aperture plane, an examination of the correlation matrix for the aperture function is helpful. The correlation matrix for the aperture function in any coordinate system is:

    $F_{ij} = \oint r_{i}(\theta,s)\, r_{j}(\theta,s) df/\oint df$
    
\noindent where $f$ is the distance along the aperture edge from the $s=0$ point.

Since this matrix is positive semi-definite, it has three real orthonormal eigenvectors $\nvec{n}_{1}$,   $\nvec{n}_{2}$, and  $\nvec{n}_{3}$ and three corresponding real, positive eigenvalues $\lambda_{1}$, $\lambda_{2}$, and $\lambda_{3}$. There exists a rotation matrix which will diagonalize this matrix, with the three positive eigenvalues on the diagonal, and, assuming the variation of $\vect{f}[s]$ normal to the aperture plane is less than the variation of $\vect{f}[s]$ in any direction in the aperture plane, the eigenvector with the smallest eigenvalue will specify the axis perpendicular to the aperture plane. We can specify this as the  $\nvec{n}_{3}$ vector. The other two eigenvectors could be chosen as an aperture-defined $x$ and $y$ axis in the aperture plane, as long as those eigenvalues are not equal to each other. If they are equal, then any linear combination of the two will also be an eigenvector of the correlation matrix, and no unique $x$ and $y$ axis may be defined from the correlation matrix. This is only a problem for a circular aperture. For example, a square aperture will have the two eigenvalues equal, yet one corner of the square could serve as a constant point of orientation. There are actually three pairs of unit eigenvectors, each element of a pair being opposite in sign. The organism has a sense of forward direction, and essentially knows the proper sign of  $\nvec{n}_{3}$, and it will be such that $\nvec{n}_{3} \cdot \vect{v}\,>\,0$. The question of how the signs of the other two eigenvectors is determined is unresolved. This means that although we do not know how the organism stores the $x$ and $y$ components of an aperture representation, the information described by the present model's choice of aperture system is nevertheless stored.

Once the aperture coordinate system has been defined, we need to find its relationship, at $\theta=0$, to our chosen global coordinate system. Except for position vectors, the two will generally be related by a rotation, which requires three independent parameters. Cortie specifies this rotation by introducing three global Euler-like angles  $\phi$, $\mu$ and $\Omega$ as parameters of the model. The present model specifies the rotation by realizing that the  $\vect{u}$ vector in the aperture system coincides with a fixed vector in the global system, , and that the local  $\vect{v}$ parameter coincides with the global direction of the centerline,  $d\vect{r}_{c}[\theta]/dt$. The relationship between systems is thus specified by the local $\vect{u}$ and $\vect{v}$ vectors and the choice of local and global coordinate system, without requiring three new parameters.

In Cortie's model, the growth axis ($D\nvec{u}$) is specified to be the global $-z$ axis, (i.e. $\nvec{u}=-\nvec{z}$) and the aperture origin is the average of the aperture function for most cases (except Lyria and Epitonium). The aperture plane at $\theta=0$ is taken to be the global $xz$ plane, rather than the $xy$ plane as in the present model. For example, in the case of an ellipse, Cortie's aperture function contains any rotation of the ellipse away from the $xz$ plane at time zero. The rotation requires three parameters and Cortie's model uses the $\phi$, $\mu$ and $\Omega$ parameters to specify such a rotation. 

\section{Conclusion}

A physical model has been developed for the growth of self-similar seashells, expressed in differential form by Eq. \ref{diffeq} and in an integrated form by Eq. \ref{inteq}. As a physical model, it is expressed in coordinate-free notation involving length and time. The model may be expressed in terms of a chosen global coordinate system, at rest with respect to the shell, and using measured parameters of the shell, or in an aperture coordinate system fixed to the moving aperture, and using local parameters, which are the parameters used by the organism to build the shell. The present model makes specific choices of these coordinate systems, as does the model of Cortie\cite{Cortie}. Cortie's model can be expressed in terms of the present model, and the measurements of Cortie on various species of mollusk can be used to calculate the local parameters of the present model. In Appendix II below,  \textit{Mathematica} code has been provided which implements transformations between coordinate systems, and plots various shells in both the local and global versions of the present model.

 Hopefully, this paper will provide both a general mathematical framework by which various models may be compared, and an approach that will be helpful in determining the information  needed by the organism in other higher-order models of the shell.

\section{Appendix I - The Frenet Coordinate System}

As developed by Illert\cite{Illert1989} and used by Fowler\cite{Fowler}, the Frenet coordinate system is sometimes used as a local coordinate system. The Frenet coordinate system is based on the centerline velocity vector $\vect{v}$ and the centerline acceleration vector $\vect{a}$. At time $\theta=0$:

    $\vect{a}=\frac{d^2 \vect{r}_{c}}{d \theta^2} = \Tens{S}\cdot\vect{v}$

\noindent and as a function of time, $\vect{a}[\theta]=e^{\tens{S}\theta}\vect{a}$.

The disadvantage of the Frenet coordinate system is that it cannot be used by the organism, since the  aperture coordinate system, defined by the aperture, is the only coordinate system available to it. The organism cannot store a non-trivial representation of a vector parameter in a coordinate system defined that vector. Additionally, the $\vect{v}$ and $\vect{a}$ vectors are not sufficient to determine the handedness of the shell, so that the $D$ parameter cannot be derived from them, and must be defined \textit{a priori} \footnote{Illert\cite{Illert1989} allows for negative values of $\alpha$ to determine $D$ which is equivalent to using $D\alpha$ in this paper in which $\alpha$ is restricted by $0\le\alpha\le\pi/2$. }

The advantage of the Frenet coordinate system is that the representations of many of the vectors and tensors involved in the description of the shell take a particularly simple form, which simplifies the mathematical expression of a model of the shell. It also provides a simple system to deal with the inclination of the aperture plane with respect to the velocity vector, which Illert\cite{Illert1989} separates into prosoclinal, orthoclinal, and (non-physical) opisthoclinal orientations according to whether the component of the normal vector $\nvec{z}$ in the $\vect{v}\,\vect{a}$ plane lies between, on or outside those two vectors, respectively.

Defining $v=|\vect{v}|$ and $a=|\vect{a}|$ the coordinate vectors of the Frenet coordinate system form a right-handed coordinate system and are defined as:
    
     \begin{math}
     \begin{array}{l}
     \nvec{e}_z =\vect{v}/v\\
     \nvec{e}_y =\vect{v}\times\vect{a}/|\vect{v}\times\vect{a}|\\
     \nvec{e}_x =\nvec{e}_{y}\times\nvec{e}_{z}
     \end{array} 
     \end{math}

\noindent In the aperture coordinate system, the basis unit vectors are represented as:

     \begin{math}
     \begin{array}{l}
     \nvec{x} = (1,0,0)\\
     \nvec{y} = (0,1,0)\\
     \nvec{z} = (0,0,1)
     \end{array} 
     \end{math}

\noindent which allows the transformation matrix between the aperture coordinate system and Frenet coordinate systems to be calculated:

    $\Tens{F}= $ \begin{math} \left(
    \begin{array}{ccc}
    -\sin[\alpha]                           & \cos[\alpha]                            & 0                       \\
    -D\cos[\alpha]  \cos[\gamma] & -D \sin[\alpha]  \cos[\gamma]  & \sin[\gamma]    \\
    \cos[\alpha]\sin[\gamma]       & \sin[\alpha]\sin[\gamma]        & D \cos[\gamma]  
    \end{array} \right) \end{math}

\noindent where for any vector $\vect{q}$, $\vect{q}_{A}=\Tens{F}\cdot\vect{q}_{F}$ and the transpose of $\Tens{F}$ is its inverse. In terms of $\vect{v}$ and  $\vect{a}$ it can be shown that:

    $\cot[\alpha] =(\vect{v}\cdot\vect{a})/v^2$
    
\noindent and:
    
    $\sin[\gamma] = |\vect{v} \times \vect{a}|/v^2$
    
\noindent The following representations in the Frenet coordinate system (subscript $F$)  result:

    $\nvec{u}_{F} = (0,\sin[\gamma],D\cos[\gamma])$
    
    $\vect{v}_{F}=(0,0,v)$
    
    $\vect{a}_{F}=v (\sin[\gamma],0,\cot[\alpha])$
    
    $\vect{r}_{cF}[0]= v $ \begin{math} \left(
    \begin{array}{c}
    -\sin[\alpha]^2 \sin[\gamma] \\
    D \sin[\alpha]^2 \tan[\alpha]\cos[\gamma]  \sin[\gamma] \\
    \tan[\alpha] (\cos[\alpha]^2 + \sin[\alpha]^2\cos[\gamma]^2 )
    \end{array} \right) \end{math}
    
    $A =v\,\tan[\alpha]\sqrt{1-\sin[\alpha]^2\sin[\gamma]^2}$

\noindent The Frenet representations for $\Tens{S}$ and $\Tens{T}$ follow directly. The normal to the aperture plane is given by:

    $\nvec{z}_{F} = (\cos[\alpha]  \sin[\gamma], \,\sin[\alpha]  \sin[\gamma] ,\, D \cos[\gamma)$
    
\section{Appendix II - \textit{Mathematica} Code}

The following modules are written for clarity rather than efficiency. There are three datasets, "Global" for the global parameters of the present model, "Local" for the local parameters in the aperture frame, and "CortieGlobal" for the global parameters of Cortie.

\subsection{Auxiliary Modules}
 
\begin{verbatim}
(* Calculate the present model transformation matrix Ro given the \
local and global values of u and v *)
Clear[GetRo1]
GetRo1[uG_, uL_, vG_, vL_] := 
 Module[{k, K, PL, PG, nPL, nPG, Cos\[Epsilon], Sin\[Epsilon], Ro},
  k = Cross[vL - vG, uL - uG];
  k = {kx, ky, kz} = k/Sqrt[k.k]; (* the axis of rotation for Ro *)
  
  K = {{0, -kz, ky}, {kz, 0, -kx}, {-ky, kx, 0}}; (* 
  The generator for Ro *)
  PL = -K.K.uL; (* 
  The projection of uL onto the plane of k (assumes uL.uL\[Equal]1) *)

    PG = -K.K.uG;  (* 
  The projection of uG onto the plane of k (assumes uG.uG\[Equal]1) *)

    If[Chop[PL.PL] == Chop[PG.PG], 
   Null, {Print ["GetRo1 ERROR1: PL.PL != PG.PG"], Abort[]}, 
   Print["GetRo1 WARNING: PL.PL==PG.PG cannot be verified"]];
  nPL = PL/Sqrt[PL.PL];(* normalized projection of uL *)
  
  nPG = PG/Sqrt[PG.PG];(* normalized projection of uG *)
  
  Cos\[Epsilon] = nPL.nPG; (* Cosine of the angle of rotation *)
  
  Sin\[Epsilon] = k.Cross[nPL, nPG];  (* 
  Sine of the angle of rotation *)
  
  Ro = IdentityMatrix[3] + 
    Sin\[Epsilon] K + (1 - Cos\[Epsilon]) K.K; (* 
  The rotation (Rodriguez formula) *)
  
  If[Chop[Ro.uL - uG] != 0, Print["GetRo1 ERROR 2"]];
  If[Chop[Ro.vL - vG] != 0, Print["GetRo1 ERROR 3"]];
  Ro
  ]
\end{verbatim}

\begin{verbatim}
(* Calculate the present model transformation matrix Ro given \[Mu], \
\[Phi], and \[CapitalOmega] *)
Clear[GetRo2];
GetRo2[\[Mu]_, \[Phi]_, \[CapitalOmega]_] := 
 Module[{g1, g2, ig2, RoC, RoC\[CapitalOmega], RoC\[Mu], RoC\[Phi], 
   Ro},
  g1 = {{1, 0, 0}, {0, 1, 0}, {0, 0, -dd}};
  g2 = {{1, 0, 0}, {0, 0, 1}, {0, 1, 0}};
  ig2 = Inverse[g2];
  RoC\[CapitalOmega] = {{Cos[\[CapitalOmega]], Sin[\[CapitalOmega]], 
     0}, {-Sin[\[CapitalOmega]], Cos[\[CapitalOmega]], 0}, {0, 0, 
     1}};
  RoC\[Mu] = {{1, 0, 0}, {0, Cos[\[Mu]], -Sin[\[Mu]]}, {0, Sin[\[Mu]],
      Cos[\[Mu]]}};
  RoC\[Phi] = {{Cos[\[Phi]], 0, -Sin[\[Phi]]}, {0, 1, 
     0}, {Sin[\[Phi]], 0, Cos[\[Phi]]}};
  RoC = RoC\[CapitalOmega].RoC\[Mu].RoC\[Phi];
  Ro = g1.RoC.ig2;
  Ro
  ]
\end{verbatim}
  
\begin{verbatim}
(* Natalina Cafra from Cortie *)
SetNatalina[] := Module[{},
  \[Alpha] = 80 Degree;
  \[Beta] = 40 Degree;
  \[Phi] = 55 Degree;
  \[Mu] = 30 Degree;
  \[CapitalOmega] = 10 Degree;
  dd = 1;
  aa = 25;
  a = 12;
  b = 16;
  uG = {0, 0, 1};
  ]
\end{verbatim}

\subsection{Global to Local Parameter Transformation}

The \textit{Mathematica} module below calculates the local aperture parameters   $\alpha$, $\vect{u}_{A}$ $\vect{v}_{A}$ and $\vect{f}_{A}$ from the global parameters $\alpha$, $\beta$, $\phi$, $\mu$, $\Omega$, $D$ (as dd), $A$ (as aa) and $\vect{f}_{P}$, according to the transformation matrix $\Tens{R}_{o}$.

\begin{verbatim}
Clear[GlobalToLocal];
GlobalToLocal[{\[Alpha]_, \[Beta]_, aa_, dd_, uG_, fG_, Ro_}] :=
 
 Module[{uGx, uGy, uGz, UG, II, SG, ro, uL, vL, fL, vG, vGx, vGy,
   vGz, iRo},
  {uGx, uGy, uGz} = uG;
  UG = {{0, -uGz, uGy}, {uGz, 0, -uGx}, {-uGy, uGx, 0}};
  II = IdentityMatrix[3];
  SG = Cot[\[Alpha]] II + UG;
  ro = aa { Sin[\[Beta]], 0, dd Cos[\[Beta]] };
  vG = {vGx, vGy, vGz} = SG.ro;
  iRo = Inverse[Ro];
  uL = iRo.uG;
  vL = iRo.vG;
  fL = iRo.fG;
  {\[Alpha], uL, vL, fL, Ro}
  ]

Test the above with Cortie's Natalina Cafra data  Natalina Cafra

SetNatalina[] (* Load Cortie's Natalina Cafra parameters *)

(* Convert local FL to global manually, and apply to GlobalToLocal *)

Rs = 1/Sqrt[Cos[s]^2/a^2 + Sin[s]^2/b^2];
ns = {Cos[s], Sin[s], 0};
Ro = GetRo2[\[Mu], \[Phi], \[CapitalOmega]];
fG = Rs Ro.ns;

Global = N[{\[Alpha], \[Beta], aa, dd, uG, fG, Ro}];
Local = {\[Alpha], uL, vL, fL, Ro} = GlobalToLocal[Global];

Print["uL = ", uL] 
Print["vL = ", vL]  
Print["fL = ", Chop[FullSimplify[fL]]] 

(* Results:
uL = {-0.709406,-0.496732,-0.5}
vL = {-9.07886,-6.35709,12.443}
fL = {(1. Cos[s])/Sqrt[0.00542535 +0.0015191 Cos[2 s]],(1. Sin[s])/Sqrt[0.00542535 +0.0015191 Cos[2 s]],0}
*)
\end{verbatim}

\subsection{Local to Global Parameter Transformation}

The \textit{Mathematica} module below calculates the global  parameters $\alpha$, $\beta$, $A$ (as aa), $D$ (as dd), $\nvec{r}_{P}$ and $\vect{f}_{P}$, from local parameters $\alpha$, $\vect{u}_{A}$ $\vect{v}_{A}$ and $\vect{f}_{A}$ according to the transformation matrix $\Tens{R}_{o}$.

\begin{verbatim}
In[89]:= Clear[LocalToGlobal]
LocalToGlobal[{\[Alpha]_, uL_, vL_, fL_, Ro_}] := 
 Module[{ux, uy, uz, UL, II, dd, Cos\[Gamma], Sin\[Gamma], Cos\[Beta], 
   Sin\[Beta], \[Beta], SL, TL, roL, aa, uG, 
   fG, \[Phi], \[Mu], \[CapitalOmega]},
  {ux, uy, uz} = uL;
  UL = {{0, -uz, uy}, {uz, 0, -ux}, {-uy, ux, 0}};
  II = IdentityMatrix[3];
  dd = uL.vL/Sqrt[(uL.vL)^2];
  Cos\[Gamma] = dd uL.vL/Sqrt[vL.vL];
  Sin\[Gamma] = Sqrt[1 - Cos\[Gamma]^2];
  Cos\[Beta] = Cos\[Gamma]/Sqrt[
   Cos[\[Alpha]]^2 + Cos\[Gamma]^2 Sin[\[Alpha]]^2];
  Sin\[Beta] = (Cos[\[Alpha]] Sin\[Gamma])/Sqrt[
   Cos[\[Alpha]]^2 + Cos\[Gamma]^2 Sin[\[Alpha]]^2];
  \[Beta] = ArcTan[Cos\[Beta], Sin\[Beta]];
  SL = Cot[\[Alpha]] II + UL;
  TL = Inverse[SL];
  roL = TL.vL;
  aa = Sqrt[roL.roL];(* also aa= Sqrt[vL.vL]Sin[\[Alpha]]Sqrt[ 1+
  Cos\[Gamma]^2 Tan[\[Alpha]]^2] *)
  uG = Ro.uL;
  fG = Ro.fL;
  
  {\[Alpha], \[Beta], aa, dd, uG, fG, Ro}
  ]

Test the above with Cortie's Natalina Cafra data  Natalina Cafra

In[78]:= SetNatalina[] (* Load Cortie's Natalina Cafra parameters *)

(* Calculate fG manually *)
Rs = 1/Sqrt[Cos[s]^2/a^2 + Sin[s]^2/b^2];
ns = {Cos[s], Sin[s], 0};
NRo = GetRo2[\[Mu], \[Phi], \[CapitalOmega]];
fG = Rs NRo.ns;

In[94]:= (* Recover Cortie's Natalina Cafra parameters *)

Global = N[{\[Alpha], \[Beta], aa, dd, uG, fG, NRo}];
Local = {\[Alpha], uL, vL, fL, NRo} = GlobalToLocal[Global];
Global2 = {\[Alpha], \[Beta], aa, dd, uG, fG, NRo} = LocalToGlobal[Local];

In[100]:= Chop[FullSimplify[Global2 - Global]]

\end{verbatim}

\subsection{Plot using Global Parameters}

The \textit{Mathematica} module below plots the shell using global parameters, including the transformation angles $\mu$, $\phi$, and $\Omega$.

\begin{verbatim}
Clear[PlotShellGlobal]
PlotShellGlobal[{\[Alpha]_, \[Beta]_, \[Phi]_, \[Mu]_, \
\[CapitalOmega]_, dd_, aa_, uG_, fG_}] :=
 
 Module[{\[Theta], uGx, uGy, uGz, UG, SG, TG, ro, Rs, vG, g1, g2, 
   ig2,
   RoC\[Phi], RoC\[Mu], RoC\[CapitalOmega], Ro, r0, r0c, tmax, tmin, 
   plot1, plot2},
  
  {uGx, uGy, uGz} = uG;
  UG = {{0, -uGz, uGy}, {uGz, 0, -uGx}, {-uGy, uGx, 0}};
  SG = Cot[\[Alpha]] II + UG;
  TG = Inverse[SG];
  ro = aa { Sin[\[Beta]], 0, dd Cos[\[Beta]] };
  vG = {vGx, vGy, vGz} = SG.ro;
  Ro = GetRo2[\[Mu], \[Phi], \[CapitalOmega]];
  r0 = Chop[N[MatrixExp[SG \[Theta]].(TG.vG + fG)]];
  r0c = Chop[N[MatrixExp[SG \[Theta]].(TG.vG + .01  fG)]];
  tmin = -8 \[Pi];
  tmax = 0;
  plot1 = 
   ParametricPlot3D[r0, {\[Theta], tmin, tmax}, {s, 0, 2 \[Pi]}, 
    PlotStyle -> Opacity[0.6], Mesh -> None, PlotPoints -> 80];
  plot2 = 
   ParametricPlot3D[r0c, {\[Theta], tmin, tmax}, {s, 0, 2 \[Pi]}, 
    PlotPoints -> 80];
  Show[plot1, plot2, AxesLabel -> {x, y, z}, PlotRange -> All]
  ]

SetNatalina[] (* Load Cortie's Natalina Cafra parameters *)

ns = {Cos[s], Sin[s], 0};
Rs = 1/Sqrt[Cos[s]^2/a^2 + Sin[s]^2/b^2];
Clear[fL, fG];
fL = Rs ns;
fG = GetRo2[\[Mu], \[Phi], \[CapitalOmega]].fL;

Global = N[{\[Alpha], \[Beta], \[Phi], \[Mu], \[CapitalOmega], dd, aa,
     uG, fG}];
PlotShellGlobal[Global]
\end{verbatim}

\begin{figure}[H]
\includegraphics[width=0.8\textwidth]{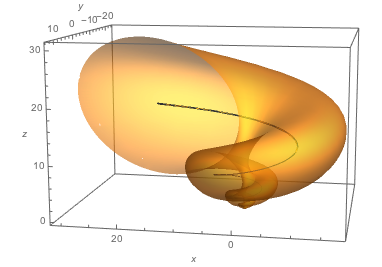}
\end{figure}

\subsection{Plot Using Local Parameters}

The \textit{Mathematica} module below plots the shell using local parameters. The $\Tens{R}_{o}$ transformation matrix is not needed to convert $\vect{f}_{A}$ and is not used, it is only kept for compatibility with the other modules. Note that the origin is the center of the aperture at time zero.

\begin{verbatim}
Clear[PlotShellLocal]
PlotShellLocal[{\[Alpha]_, uL_, vL_, fL_, Ro_}] := 
 Module[{\[Theta], UL, uLx, uLy, uLz, SL, TL, ro, r, rc, tmin, tmax, 
   plot1, plot2},
  {uLx, uLy, uLz} = uL;
  UL = {{0, -uLz, uLy}, {uLz, 0, -uLx}, {-uLy, uLx, 0}};
  II = IdentityMatrix[3];
  SL = Cot[\[Alpha]] II + UL;
  TL = Inverse[SL];
  ro = TL.vL;
  r = Chop[N[Simplify[MatrixExp[SL \[Theta]].(ro + fL) - ro]]];  (* 
  The shell *)
  
  rc = Chop[
    N[Simplify[MatrixExp[SL \[Theta]].(ro + .01  fL) - ro]]]; (* 
  The centerline, 0.01 thick *)
  tmin = -8 \[Pi];
  tmax = 0;
  plot1 = 
   ParametricPlot3D[r, {\[Theta], tmin, tmax}, {s, 0, 2 \[Pi]}, 
    PlotStyle -> Opacity[0.6], Mesh -> None, PlotPoints -> 80];
  plot2 = 
   ParametricPlot3D[rc, {\[Theta], tmin, tmax}, {s, 0, 2 \[Pi]}, 
    PlotPoints -> 80];
  Show[plot1, plot2, AxesLabel -> {x, y, z}, PlotRange -> All]
  ]

SetNatalina[] (* Load Cortie's Natalina Cafra parameters *)

(* Calculate fG from Cortie's data *)
ns = {Cos[s], Sin[s], 0};
Rs = 1/Sqrt[Cos[s]^2/a^2 + Sin[s]^2/b^2];
Ro = GetRo2[\[Mu], \[Phi], \[CapitalOmega]];
Clear[fL, fG]
fL = Rs ns;
fG = Ro.fL;

Global = N[{\[Alpha], \[Beta], aa, dd, uG, fG, Ro}];
Local = FullSimplify[GlobalToLocal[Global]];
plot1 = PlotShellLocal[Local]
\end{verbatim}

\begin{figure}[H]
\includegraphics[width=0.8\textwidth]{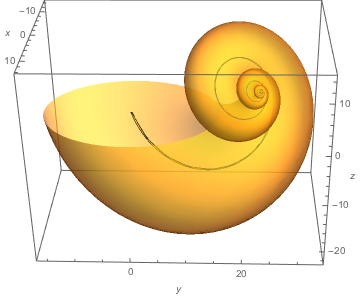}
\end{figure}

\section{Bibliography}

\end{document}